\documentclass{article}
\usepackage{spconf,amsmath,graphicx}
\usepackage{booktabs}
\usepackage{multirow}
\usepackage{siunitx}
\usepackage{ragged2e}

\title{Detecting Emotion Primitives from Speech and their use in discerning Categorical Emotions}
\name{V. Kowtha, V. Mitra, C. Bartels, E. Marchi, S. Booker, W. Caruso, S. Kajarekar, D. Naik}
\vspace{-3mm}
\address{Apple Inc., Cupertino, CA, USA}
\vspace{-3mm}
\begin{document}
%
\maketitle
\begin{abstract}
Emotion plays an essential role in human-to-human communication, enabling us to convey feelings such as happiness, frustration, and sincerity. While modern speech technologies rely heavily on speech recognition and natural language understanding for speech content understanding, the investigation of vocal expression is increasingly gaining attention. Key considerations for building robust emotion models include characterizing and improving the extent to which a model, given its training data distribution, is able to generalize to unseen data conditions. This work investigated a long-shot-term memory (LSTM) network and a time convolution - LSTM (TC-LSTM) to detect primitive emotion attributes such as valence, arousal, and dominance, from speech. It was observed that training with multiple datasets and using robust features improved the concordance correlation coefficient (CCC) for valence, by 30\% with respect to the baseline system. Additionally, this work investigated how emotion primitives can be used to detect categorical emotions such as happiness, disgust, contempt, anger, and surprise from neutral speech, and results indicated that arousal, followed by dominance was a better detector of such emotions.
\end{abstract}
\vspace{-1mm}
\vspace{-1mm}
\section{Introduction}
\label{sec:intro}
\vspace{-1mm}
Detection of intent from a query is the principal task performed by the majority of voice operated assistants. Intent is primarily detected from words which are recognized from speech, via an automated speech recognition system. Often, voice operated assistants need to disambiguate between requests that vary in terms of vocal expression, such as an urgent query versus a casual observation. In these cases, the text recognized from the speech signal may not contain sufficient information to accurately infer the user’s intent. While the traditional technological solution to this problem is to perform the disambiguation with the aid of a follow-up query, human beings are notably adept at interpreting spoken intent directly, by attending to vocal expression. Thus, this study investigates the detection of vocal expression in the form of emotion primitives through machine learning approaches. 

Detection of vocal expressions in the form of emotion has received much attention in speech technology research within the past several years, where studies have focused on devising robust and relevant acoustic features \cite{eyben2013opensmile, eyben2016gemaps}, modeling techniques \cite{parthasarathy2019ladder, Mao2014LearningSF}, multi-task learning \cite{Lotfian2018PredictingCE, Xia2017AML}, multi-modal fusion \cite{Busso2004facial, Kapoor2005multimodal}. Early studies on speech-based emotion detection have used acted or elicited emotion datasets \cite{Busso2008IEMOCAP, savee2011}  (where actors were recorded while speaking with specified emotions). Observations from acted-emotion studies revealed that models trained with acted emotions may not generalize well to spontaneous subtle emotions \cite{Cowie2005complexity} as a consequence, datasets containing spontaneous emotions were collected, such as the MSP Podcast dataset \cite{Mariooryad2014MSPpodcast}. The downside of collecting spontaneous speech emotion datasets is that they typically lack ground-truth labels, and thus require manual annotation, that suffer from varying degrees of grader agreement. Labels in emotion datasets contain either categorical emotions (such as happy, sad, neutral etc.) and/or primitive emotions (such as valence, arousal, dominance etc.). While categorical emotions are easy to interpret, they are difficult to annotate, as they often lead to annotator disagreements, skewed datasets and suffer from ambiguity in defining the lexicon for emotion categories \cite{mitra2019interspeech}. On the other hand, primitive emotions, which are defined by the valence-arousal-dominance scale, are easier to annotate but harder to interpret, and typically generate results that are more easily comparable. Moreover, primitive emotions can be coarsely aggregated, to recover categorical emotions \cite{grimm2006combining}.

In this work we used spontaneous speech corpus labelled with primitive emotions and investigated the following:
\newline \textit{(1) Role of annotator agreement on model performance.
\newline (2) Role of acoustic features on model robustness.
\newline (3) Whether such models are useful in detecting vocal expression in the form of happiness, disgust, contempt, anger, and surprise.}

Through our work we demonstrate that:
\newline \textit{(a) Use of low-dimensional frame-level features (such as filterbank energies) can demonstrate performance comparable to complex feature sets investigated in the literature.
\newline (b) Presence of additional data resources can help to improve an emotion detection model's performance and generalization capacity.
\newline (c) Simple score level fusion of multiple complimentary systems can improve the overall performance}

The outline of the paper is as follows: section (2) will present the datasets used in our study, (3) will introduce the acoustic features investigated in this work, (4) will detail the acoustic model and its parameters, in (5) we will present the results, followed by conclusion in (6).
\vspace{-1mm}
\section{Data}
\label{sec:format}
\vspace{-1mm}
We use a slightly expanded version of the data used in our earlier study \cite{mitra2019interspeech} which contains 120 hours speech material spoken in US English. The data had no speaker level information. The duration of each utterance varied between 2 to 6 seconds. The data contained perceptually assigned valence, arousal and dominance scores. For more information on this dataset, please refer to \cite{mitra2019interspeech}. Additionally, this study also uses the MSP-Podcast data \cite{Mariooryad2014MSPpodcast} that contains speech spoken by English speakers collected from online audio shows, covering topics such as politics, sports, entertainment, etc. The speech segments in this dataset contain single speaker utterances with duration between 2.75 and 11 seconds. Overall the MSP-Podcast (ver-3) data contained a little over 50 hours of speech. The data came with speaker and gender labels, which were not used in this study. There were altogether 588 speaker labels, where 50 speakers were present in the test set. The remaining data was used to train the model, where 90\% of the data was used as the training set and the remaining 10\% as the cross-validation set. We will denote the MSP-Podcast training data as MSP-train. To make our results comparable to the literature, we will report results on MSP-Podcast (version 2) eval set ($\approx$ 12 hours of speech), which we will denote as MSP-eval.
\vspace{-1mm}
\section{Acoustic features}
\label{sec:pagestyle}
\vspace{-1mm}
We investigated multiple acoustic features to parameterize speech. The baseline feature is the 40-dimensional mel-filterbank energy ($MFB$) features, appended by pitch, pitch-delta and voicing features, which we denote as $MFB+F0$ feature. We explored 40-D gammatone filter-bank energies ($GFB$) and speech modulation energies (extracted through the amplitude modulation ($MOD$) feature extraction setup as specified in \cite{mitra2012modulation}), both were appended with 3-D pitch ($F0$) and voicing features and we denote them as $GFB+F0$ and $MOD+F0$ features. We have used articulatory features in the form of vocal-tract constriction variables ($TV$) \cite{mitra2017hybrid}. In our earlier work \cite{mitra2019interspeech} we have shown that $TV$s can assist in detecting valence from the speech signal. The $TV$s define degree and location of constriction actions within the human vocal tract and have eight dimensions \cite{mitra2017hybrid, mitra2010retrieving}. Similar to \cite{mitra2019interspeech} we have used an LSTM-based speech-inversion system which takes in spliced (window of 5 frames on both sides of the current frame) $MFB+F0$ features as input and maps that to the 8 TV trajectories. 
\vspace{-1mm}
\section{Acoustic Modeling}
\label{sec:acousticmodeling}
\vspace{-1mm}
We have used single-layer LSTM networks consisting of 128 neurons in the recurrent and the embedding layers, to train the baseline primitive emotion detection (regression) model. The input to the model was low-level features described in section 4, which were analyzed with a window size of 25 ms and a frame rate of 10 ms, and the output was 3-D primitive emotions: valence, arousal and dominance. The model was tuned using a held-out dev set, and based upon that the number of neurons in each layer and the cost function (concordance-correlation-coefficient, $CCC_{cost}$ as shown in (1)) was selected. The ${CCC_{cost}}$ is a combination ($\alpha=1/3$ and $\beta=1/3$) of CCC's obtained from each of the valence, dominance and arousal dimensions. CCC for each dimension is defined by (2), where where ${\mu _{x}}$ and ${\mu _{y}}$ are the means, ${\sigma _{x}^{2}}$ and ${\sigma _{y}^{2}}$ are the corresponding variances for the estimated and ground truth variables and ${\rho}$ is the correlation coefficient between those two variables. The models were trained with a mini-batch size of 512, using Adam optimizer, with a learning rate of 0.001. For all the model training steps, early stopping was allowed based on cross-validation error. 
\vspace{-3mm}
\begin{equation}
\begin{aligned}
{CCC_{cost}:= \alpha CCC_{val}+\beta CCC_{aro}+(1-\alpha-\beta)CCC_{dom}} \\
\end{aligned}
\end{equation}
\vspace{-3mm}
\begin{equation}
\begin{aligned}
CCC &= \frac {2\rho \sigma_x \sigma_y}{\sigma_x^2+\sigma_y^2 +(\mu_x-\mu_y)^2 }
\end{aligned}
\end{equation}

\begin{figure}[htb]
\begin{minipage}[b]{1.0\linewidth}
  \centering
  \centerline{\includegraphics[width=6cm]{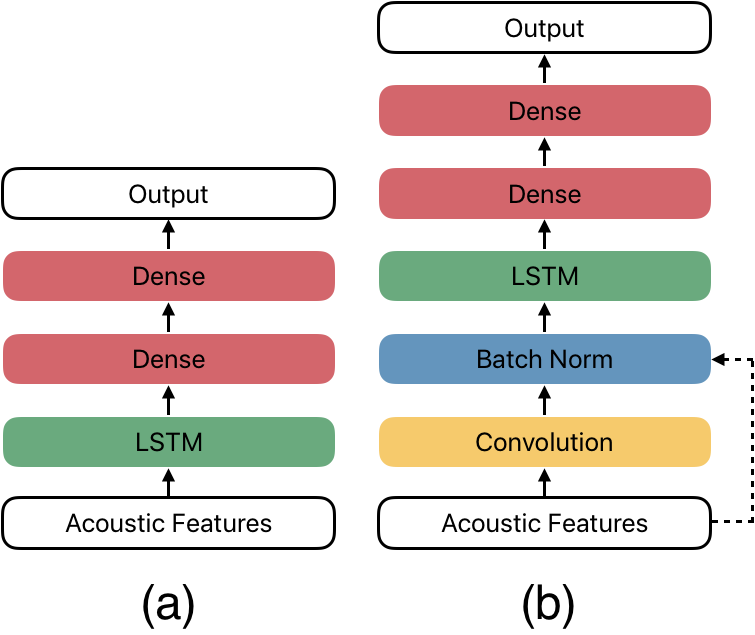}}
\end{minipage}
\caption{Architecture of (a) LSTM network and (b) Time convolution LSTM network (TC-LSTM)}
\label{fig:fig1}
\end{figure}
\vspace{1mm}
Additionally, we investigated a time-convolutional layer before the LSTM layer (TC-LSTM, shown in Figure~\ref{fig:fig1}), where the number of convolutional layers (having filter size = 3) was same as the number of input feature dimensions. The network also had a skip connection, where acoustic features were also fed directly to the LSTM layer in addition to the convolutional layer outputs. The LSTM layer had 128 and 312 neurons in the recurrent and embedding layers, respectively.
\vspace{-2mm}
\section{Results and analysis}
\label{sec:results}
\vspace{-1mm}
We trained three LSTM models with (a) internal-train, (b) MSP-train and (c) internal+MSP train data, which were evaluated with internal-eval and MSP-eval data respectively. These models were trained with $MFB+F0$ features and the results are shown in table~\ref{tab:table1}.

\vspace{-3mm}
\begin{table}[th]
\centering
\caption{Primitive emotion $CCC$ from models trained with baseline MFB+F0 features}
\vspace{1mm}
\label{tab:table1}
  \begin{tabular}{lcccccc}
    \toprule
    \multirow{2}{*}{} &
      \multicolumn{3}{c}{Internal-eval} &
      \multicolumn{3}{c}{MSP-eval} \\
      & {val} & {aro} & {dom} & {val} & {aro} & {dom} \\
      \midrule
    Int-train & 0.56 & 0.69 & \textbf{0.58} & 0.13 & 0.53 & 0.40  \\
    MSP-train & 0.14 & 0.55 & 0.47 & 0.24 & 0.73 & \textbf{0.68} \\
    Both-tr & \textbf{0.57} & \textbf{0.70} & \textbf{0.58} & \textbf{0.26} & \textbf{0.75} & \textbf{0.68} \\
    \bottomrule
  \end{tabular}
\end{table}
\vspace{-5mm}
\begin{figure}[htb]
\begin{minipage}[b]{1.0\linewidth}
  \centering
  \centerline{\includegraphics[width=9cm]{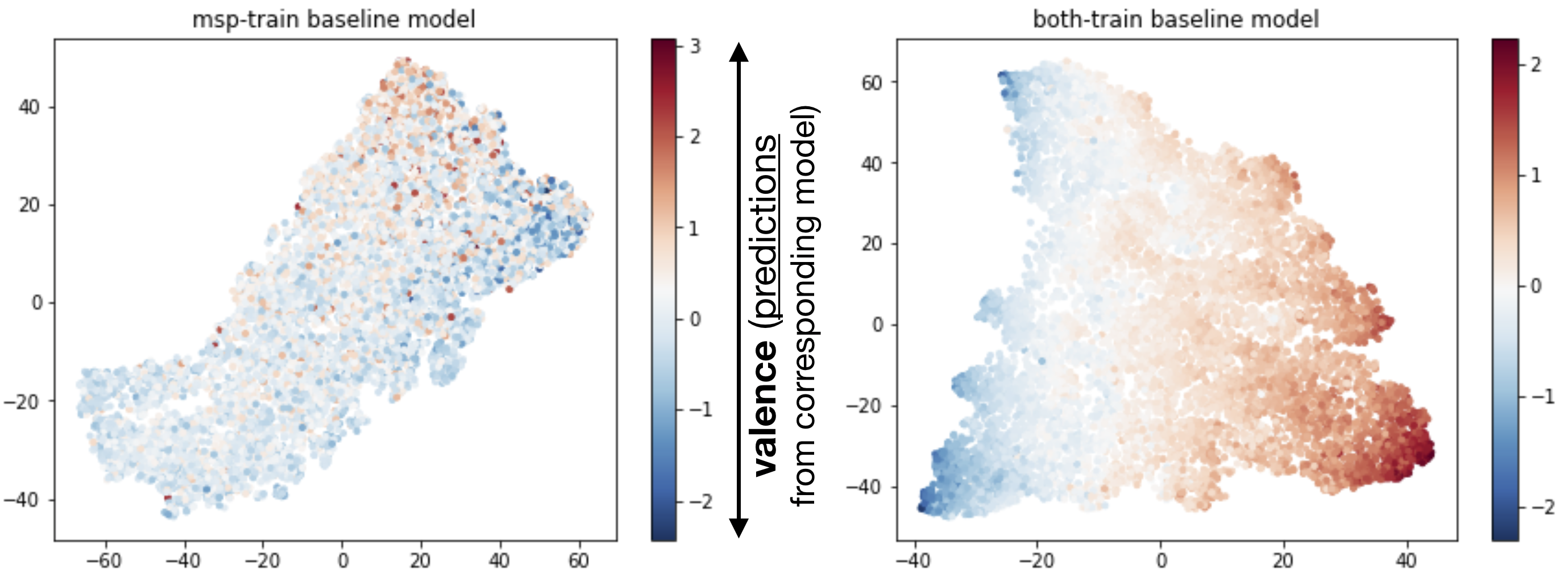}}
\end{minipage}
\caption{t-SNE plot of embeddings from MSP-train and Both-tr models, colored w.r.t valence score}
\label{fig:fig2}
\end{figure}
\vspace{-3mm}
Table~\ref{tab:table1} shows the impact of data mismatch between training and evaluation data. Model trained with the internal-train data did better on the internal-eval set, while the model trained with MSP-train, performed better for MSP-eval set, and both failing to demonstrate similar performance on the mismatched eval set (MSP-eval and internal-eval sets, respectively). This provides some realistic interpretation of how much performance gets impacted due to domain mismatch, which is found to be $\approx$ 40\% reduction in $CCC$. The last row in Table~\ref{tab:table1} shows that a model trained with internal-train and MSP-train data (followed by fine-tuning with each of those training sets) demonstrated a much better performance for both the eval sets. Given, that the Both-train model gave the best baseline $CCC$ performance for both the eval sets, we will be using that model as our baseline in the rest of this paper. Figure-\ref{fig:fig2} shows the t-SNE plot of the embeddings obtained from MSP-train and Both-train models for different values of valence, where we can observe that the latter model has a better separation of high-valence data points from low-valence ones.
\vspace{-3mm}
\subsection{Role of Annotator Consensus}
\label{ssec:graderagree}
\vspace{-1mm}
A careful analysis of the evaluation set indicated that annotation consensus played an important role on evaluation performance. The MSP-podcast data came with annotator decisions on the primary categorical emotion, where the number of annotators grading a specific utterance varied from five to sixteen. The primary categorical emotion consensus on the evaluation set is determined by the number of annotators who selected category that received the majority vote divided by the total number of annotators who voted for that utterance \cite{tang2011semi}. We denote an utterance’s consensus ($p(C)$) by the probability of an annotator selecting the primary categorical emotion that received the majority vote \cite{tang2011semi}. Table~\ref{tab:table2} presents the results from the MSP-eval set as obtained from the baseline model, when grouped by annotator consensus.

Table ~\ref{tab:table2} presents some interesting observations on the role of annotator consensus on performance evaluation. While valence $CCC$ always increased with increase in consensus (last three rows in table 2), arousal and dominance remained relatively stable. This may indicate that valence may be relatively difficult to annotate and can be correlated with the consensus emotion decision, compared to arousal and dominance. Interestingly, at extremely low consensus all of the three primitive emotions demonstrated extreme deterioration in performance, indicating low-consensus data may not be reliable in assessing a model’s performance. From these observations we can claim that:
\newline \textit{(1) Primitive emotion models are sensitive to annotator consensus, where valence was found to be more sensitive to it compared to arousal and dominance.
\newline (2) Data with less than 25\% consensus may not be suitable for assessing the goodness of a model, and it may be useful to not consider such data points.}

\vspace{-3mm}
\begin{table}[th]
\centering
\caption{Primitive emotion $CCC$ from multi-condition trained baseline model trained when MSP-eval set is grouped by annotator consensus}
\vspace{1mm}
\label{tab:table2}
  \begin{tabular}{lccc}
    \toprule
    \multirow{2}{*}{Annotator Consensus} &
      \multicolumn{3}{c}{MSP-eval ($CCC$)} \\
      & {val} & {aro} & {dom} \\
      \midrule
    $p(C)$ $\leq$ 0.25 & 0.00 & 0.59 & 0.57  \\
    $p(C)$ $\leq$ 0.40 & 0.12 & 0.73 & 0.67  \\
    $p(C)$ $\geq$ 0.50 & 0.32 & 0.77 & \textbf{0.70}  \\
    $p(C)$ $\geq$ 0.60 & 0.33 & 0.77 & 0.69  \\
    $p(C)$ $\geq$ 0.75 & 0.40 & \textbf{0.78} & \textbf{0.70}  \\
    $p(C)$ $\geq$ 0.90 & \textbf{0.46} & \textbf{0.78} & 0.69  \\
    \bottomrule
  \end{tabular}
\end{table}
\vspace{-3mm}

\begin{figure*}[htb]
\begin{minipage}[b]{1.0\linewidth}
  \centering
  \centerline{\includegraphics[width=17.5cm]{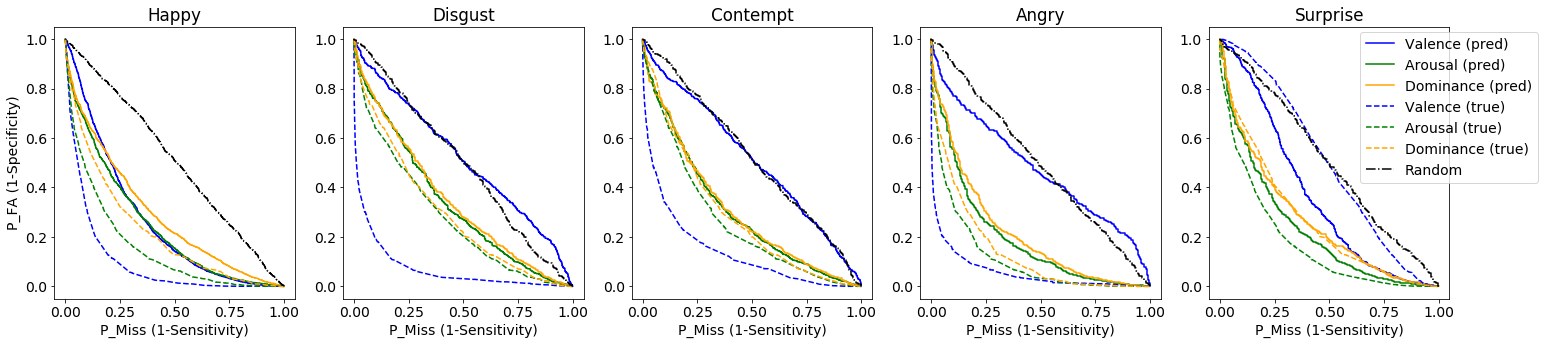}}
\end{minipage}
\caption{ROC curves for detecting categorical emotions: happy, disgust, contempt, angry and surprise from neutral.}
\label{fig:fig3}
\end{figure*}
\vspace{-5mm}
\begin{table}[th]
\centering
\caption{Primitive emotion $CCC$ from LSTM models trained with different acoustic features}
\vspace{1mm}
\label{tab:table3}
  \begin{tabular}{lcccccc}
    \toprule
    \multirow{2}{*}{} &
      \multicolumn{3}{c}{Internal-eval} &
      \multicolumn{3}{c}{MSP-eval} \\
      & {val} & {aro} & {dom} & {val} & {aro} & {dom} \\
      \midrule
    $MFB+F0$ & \textbf{0.57} & \textbf{0.70} & \textbf{0.58} & 0.26 & \textbf{0.75} & \textbf{0.68}  \\
    $GFB+F0$ & 0.56 & 0.68 & 0.57 & 0.29 & 0.72 & 0.63 \\
    $MOD+F0$ & 0.56 & 0.69 & 0.56 & \textbf{0.30} & 0.72 & 0.63 \\
    \bottomrule
  \end{tabular}
\end{table}
\vspace{-3mm}
\begin{table}[th]
\centering
\caption{Primitive emotion $CCC$ from TC-LSTM models trained with different acoustic features}
\vspace{1mm}
\label{tab:table4}
  \resizebox{\columnwidth}{!}{\begin{tabular}{lcccccc}
    \toprule
    \multirow{2}{*}{} &
      \multicolumn{3}{c}{Internal-eval} &
      \multicolumn{3}{c}{MSP-eval} \\
      & {val} & {aro} & {dom} & {val} & {aro} & {dom} \\
      \midrule
    $MFB+F0$ & 0.56 & 0.69 & 0.57 & 0.27 & \textbf{0.75} & \textbf{0.69}  \\
    $GFB+F0$ & 0.58 & \textbf{0.70} & \textbf{0.58} & 0.32 & 0.74 & 0.65 \\
    $MOD+F0$ & 0.57 & \textbf{0.70} & \textbf{0.58} & 0.31 & 0.74 & 0.67 \\
    $MOD+TV+F0$ & \textbf{0.59} & \textbf{0.70} & \textbf{0.58} & \textbf{0.33} & 0.74 & 0.68 \\
    \bottomrule
  \end{tabular}}
\end{table}
\vspace{-4mm}
\begin{table}[th]
\centering
\caption{Primitive emotion $CCC$ from TC-LSTM models and the relevant state-of-the-art}
\vspace{1mm}
\label{tab:table5}
  \resizebox{\columnwidth}{!}{\begin{tabular}{lcccccc}
    \toprule
    \multirow{2}{*}{} &
      \multicolumn{3}{c}{MSP-eval} \\
      & {val} & {aro} & {dom} & {Params} \\
      \midrule
    CNN $MFB$  \cite{parthasarathy2019ladder} & 0.25 & 0.74 & 0.66  & 800K\\
    Best in \cite{parthasarathy2019ladder}* & 0.30 & \textbf{0.77} & \textbf{0.70} & \textgreater1M\\
    TC-LSTM $MODTV+F0$ & 0.33 & 0.74 & 0.68 & 100K \\
    {TC-LSTM $MOD+TV+F0$} \\
     + {TC-LSTM $GFB+F0$} & \textbf{0.34} & \textbf{0.77} & 0.69 & 200K \\
    \bottomrule
  \end{tabular}}
\end{table}
\vspace{1mm}
\subsection{Robustness}
\label{ssec:robustness}
\vspace{-1mm}
To investigate if model performance can be improved beyond the baseline, we trained several models using $GFB+F0$ and $MOD+F0$ features. The results from that study is shown below in Table~\ref{tab:table3}. Table~\ref{tab:table3} shows that the $GFB+F0$ and $MOD+F0$ demonstrated comparable performance with respect to the baseline $MFB+F0$ features, but with improved (statistically significant) CCC for valence on the MSP-eval sets. Table~\ref{tab:table4} shows the results obtained from the TC-LSTM acoustic model, where it can be seen that the TC-LSTM overall performed better than the LSTM model. We observed that the best result from our TC-LSTM model is better than a comparable $MFB$-CNN model (multi-task learning) that was evaluated on the same MSP\_eval set (shown in Table~\ref{tab:table5}), and was very close to the best performing (* in Table~\ref{tab:table5}) system that used more than 6K feature.
\vspace{-3mm}
\subsection{Application: Detection of Categorical Emotions from their primitives}
\vspace{-1mm}
\label{ssec:categoricalemotions}
To investigate how primitive emotion decisions generalize to categorical emotions, we investigated the task of detecting happiness, disgust, contempt, anger and surprise versus neutral, given the valence, arousal and dominance scores from the $MFB+F0$ LSTM model. Figure~\ref{fig:fig3} shows the ROC curves for detecting the respective categorical emotions from neutral, given the predictions (\textit{pred}) from the $MFB+F0$ LSTM model and the \textit{true} label. Table~\ref{tab:table6} presents the area under the curve (AUC) for each of these cases and shows that while arousal (\textit{pred}) is a strong indicator for all the categories, valence is useful for detecting happiness, dominance for disgust, contempt and anger. ROC in Figure~\ref{fig:fig3} shows that the (\textit{true}) valence is the best indicator for detecting happiness, disgust, contempt and anger however (\textit{pred}) valence shows much worse performance compared to it; which motivates the necessity to improve the detection of valence from speech.
\vspace{-5mm}
\begin{table}[th]
\centering
\caption{Primitive Emotion AUC from TC-LSTM models trained with different acoustic features}
\vspace{1mm}
\label{tab:table6}
  \resizebox{\columnwidth}{!}{\begin{tabular}{lcccccc}
    \toprule
    \multirow{2}{*}{} &
      \multicolumn{5}{c}{AUC} \\
      & {Happy} & {Disgust} & {Contempt}& {Anger}& {Surprise} \\
      \midrule
    Val & 0.36 & 0.48 & 0.49& 0.52 & 0.36  \\
    Aro & \textbf{0.34} & \textbf{0.33} & \textbf{0.30}& \textbf{0.19} & \textbf{0.22} \\
    Dom & 0.39 & 0.35 & \textbf{0.30} & 0.21 & 0.26 \\
    \bottomrule
  \end{tabular}}
\end{table}
\vspace{-3mm}
\section{Conclusion}
\label{sec:conclusion}
\vspace{-1mm}
We investigated a TC-LSTM  model to detect primitive emotions from speech, and demonstrated that it performed significantly better in detecting valence compared to the state-of-the-art reported in the literature for a publicly available dataset. We observed that $MOD+TV+F0$ features offered the best performance for detecting valence, while $MFB+F0$ features performed better for detecting arousal and dominance. We demonstrated that frame-level filterbank energy features can generate comparable performance to that of large dimensional features typically used in the literature. We also observed that simple score level fusion can improve overall emotion detection performance. Finally, we observed that the model generated primitive emotion scores are useful to detect categorical emotions. 
\vspace{-3mm}
\section{Acknowledgement}
\label{sec:acknowledgement}
The authors would like to thank Russ Webb and Panayiotis Georgiou for their valuable comments and suggestions. 

\bibliographystyle{IEEEbib}
\bibliography{strings,refs}

\end{document}